\documentclass[twocolumn,showpacs,amsmath,amssymb]{revtex4}
\usepackage{graphicx}

\begin{document}

\title{VBS/CFT Correspondence and Thermal Tensor Network}
\author{Hiroaki Matsueda\footnote{matsueda@sendai-nct.ac.jp}}
\affiliation{Sendai National College of Technology, Sendai 989-3128, Japan}
\date{\today}
\begin{abstract}
It has been recently observed that the reduced density matrix of a two-dimensional (2D) valence bond solid state can be mapped onto the thermal density matrix of a 1D Heisenberg quantum spin chain. Motivated by the observation, I propose a very simple phenomenological theory for this type of correspondence based on a finite-temperature tensor network formalism recently developed. I adress close relationship and sharp difference between the present correspondence and multiscale entanglement renormalization ansatz in terms of network geometry.
\end{abstract}
\maketitle

Quantum / classical correspondence and its variants offer many interesting questions to modern physics~\cite{Matsueda}. Usually, it is easy to handle classical systems (or possibly gapped quantum systems), and thus the correspondence plays a crucial role on the analysis of, in particular, strongly interacting quantum systems. Useful measures to detect the correspondence between completely different systems are the entanglement entropy and the entanglement spectrum, since they represent the amount of information and universality irrespective of details of the models.

In condensed matter physics, the most well-known and useful correspondence is the so-called Suzuki-Trotter decomposition that is applied to quantum Monte Carlo simulations~\cite{Suzuki}. In the decomposition, a spatially $d$-dimensional ($d$D) quantum system is located at the boundary of a corresponding $(d+1)$-dimensional classical space with the additional imaginary time. After compactification by taking appropriate boudary conditions, the classical system becomes a torus in a case of $d=1$. Thus, the correspondence is closely related to the spacetime structure of our model space. On the other hand, the torus boundary condition also generates correspondence between quite different systems, where we consider a gapped quantum system instead of a classical one. It is a recent topic that the compactification due to the boundary condition of 2D gapped systems such as a valence bond solid state (VBS) leads to 1D critical systems characterized by a conformal field theory (CFT)~\cite{Cirac,Lou,Poilblanc,Poilblanc2,Tanaka,Santos,Stephan,Stephan2,Zaletel,Verstraete}. This is roughly understood by starting with a 2D tensor product state with small bond dimension representing gapped cases and by taking trace of tensors along one direction. Then, the result is a matrix product state (MPS) with large bond dimension, leading to critical behavior.

In this brief report, I would like to propose a simple phenomenological approach to this problem based on a finite-tempererature tensor network formalism that we have resently developed~\cite{Matsueda2,Matsueda3,Matsueda4}. We will find that the VBS/CFT correspondence has a tensor network structure similar to the MERA network~\cite{Vidal} except for metric on the holographic space.

Let us introduce an identity state $\left|I\right>$ and corresponding thermal state $\left|\psi(\beta)\right>$ as
\begin{eqnarray}
\left|n\right> &=& \left|n_{1},...,n_{L}\right>\otimes\left|\tilde{n}_{1},...,\tilde{n}_{L}\right> , \\
\left| I\right> &=& \sum_{\{n_{j}\}}\left|n\right> , \\
\left|\psi(\beta)\right> &=& \rho^{1/2}_{thermal}(\beta)\left| I\right> = \sum_{\{n_{j}\}}R^{n_{1}\cdots n_{L}}\left|n\right> , \label{wf} \\
R^{n_{1}\cdots n_{L}} &=& \sum_{\sigma}A_{\sigma}^{n_{1}\cdots n_{L}}A_{\sigma}^{\tilde{n}_{1}\cdots\tilde{n}_{L}} , \label{wf2}
\end{eqnarray}
where $R[=\rho^{1/2}_{thermal}(\beta)]$ is the square root of the density matrix of the total system including the tilde space. Precisely saying, they are formulated by means of thermofield dynamics~\cite{Takahashi}. We can choose any identity state $\left|I\right>$ in principle, but in the present case we assume that $\left|n\right>$ is the eigenstate of the Hamiltonian~\cite{Suzuki2}.

\begin{figure}[htbp]
\begin{center}
\includegraphics[width=7.5cm]{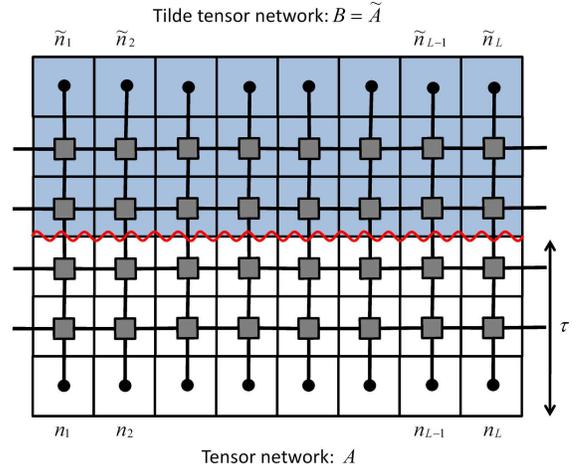}
\end{center}
\caption{Thermal tensor network. Black dots and grey squares are respectively the original sites in a 1D quantum system and tensors that come from decomposition of $R^{n_{1},...,n_{L}}$ in Eq.~(\ref{wf2}). The wavy line represents an event horizon labeled by $\sigma$, and the blue area is  the tilde tensor network.}
\label{fig1}
\end{figure}

Figure~\ref{fig1} is a typical thermal tensor network for 1D quantum systems. The central charge is taken to be $c$. Because of the presence of the tilde tensor network, the original tensor network is doubled, and they are pasted together by the index $\sigma$ as shown in Fig.~\ref{fig1}. The boundary between them becomes an event horizon. When we consider the MERA network under an appropriate metric, the layer number of the network is proportional to the inverse temperature in the 1D system~\cite{Matsueda2}. Instead of the curved metric, here we assume that the tensor aligns on the 2D square lattice: $\sigma=\{m_{1},m_{2},...,m_{L}\}$. Clearly, the total system is defined on a uniform 2D square lattice, since both of the original and the tilde Hilbert spaces are identical in definition. The tensor decomposition yields that the dimension of each tensor is not so large even if we start from a 1D critical system. In connection with VBS/CFT, we can regard the total network as a gapped quantum state such as VBS.

\begin{figure}[htbp]
\begin{center}
\includegraphics[width=6cm]{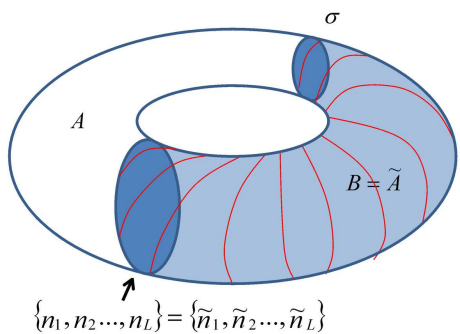}
\end{center}
\caption{Rolling the original 2D sheet into a torus with boundary conditions.}
\label{fig2}
\end{figure}

We introduce two boundary conditions to this open system in order to compactify the system: one is periodic boundary condition along the $x$ direction (we first take this condition), and the other one connects the original with the tilde Hilbert spaces, since they are maximally entangled. Eventually, the system becomes a torus as shown in Fig.~\ref{fig2}. We know that this setup is almost equivalent to Cirac's and Lou's approaches~\cite{Cirac,Lou}. In the present case, cutting the torus into two at two boundaries creates two identical tensor networks, and then the two boundaries are characterized by the horizon index $\sigma$ and the site indices $n_{1}, n_{2}, ..., n_{L}$.

The key ingredient of VBS/CFT is that the reduced density matrix of a 2D VBS state can be mapped onto the thermal density matrix of a 1D Heisenberg quantum spin chain. We prove this conjecture by the thermal tensor network. Here we take $B=\tilde{A}$, and tracing over $B$ degree of freedom can be represented as follows:
\begin{eqnarray}
{\rm Tr}_{\tilde{A}}O = \sum_{\{n_{j}\}}\left<\tilde{n}_{1},...,\tilde{n}_{L}\right|O\left|\tilde{n}_{1},...,\tilde{n}_{L}\right> ,
\end{eqnarray}
for an arbitrary operator $O$. The total Hamiltonian (hat Hamiltonian) is given by $\hat{H}=H-\tilde{H}$ and $\hat{H}\left|\psi(\beta)\right>=0$.

The proof is very simple. Let us deform the partial density matrix $\rho_{A}$ as follows:
\begin{eqnarray}
\rho_{A} &=& {\rm Tr}_{\tilde{A}}\left|\psi(\beta)\right>\left<\psi(\beta)\right| \nonumber \\
&=& \sum_{\{n_{j}\}}\left<\tilde{n}_{1},...,\tilde{n}_{L}\right|\left.\psi(\beta)\right>\left<\psi(\beta)\right.\left|\tilde{n}_{1},...,\tilde{n}_{L}\right> \\
&=& \sum_{\{n_{j}\}}R^{n_{1},...,n_{L}}\left(R^{n_{1},...,n_{L}}\right)^{\ast} \nonumber \\
&& \;\;\;\;\;\times\left|n_{1},...,n_{L}\right>\left<n_{1},...,n_{L}\right| .
\end{eqnarray}
We clearly find
\begin{eqnarray}
R^{n_{1},...,n_{L}}\left(R^{n_{1},...,n_{L}}\right)^{\ast} = \bigl|\left<n\right|\rho^{1/2}_{thermal}(\beta)\left|n\right>\bigr|^{2} . \label{equality}
\end{eqnarray}
Therefore, the entanglement entropy for the 2D tensor network state is roughly the thermal density matrix of the corresponding 1D quantum system. When the temperature is high and the quantum fluctuation can be neglected, the right hand side of Eq.~(\ref{equality}) really approaches $\rho_{thermal}$. According to my previous work on quantum-to-classical correspondence in the Ising model ($c=1/2$) and the three-state Potts model ($c=4/5$)~\cite{Matsueda5}, we need only one particular classical spin configuration $\left|n\right>$ at a critical point. Thus, if our system is critical, Eq.~(\ref{equality}) should be the thermal density matrix.

\begin{figure}[htbp]
\begin{center}
\includegraphics[width=7.5cm]{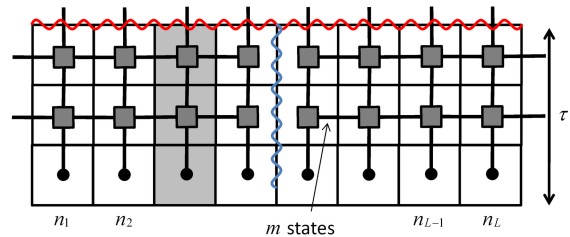}
\end{center}
\caption{Definition of $m$, $\chi$, and $\tau$. The total degree $\chi$ at the blue wavy line is given by $\chi=m^{\tau}$. Gray area represents a matrix if we describe this system by MPS.}
\label{fig3}
\end{figure}

In a viewpoint of close connection between thermal MERA network and a string-theory result, the temperature is a function of the layer number of the original tensor network as we have already mentioned~\cite{Azeyanagi,Matsueda2}. In those cases, the network has discrete hyperbolic metric. On the other hand, the metric of the present case is flat Euclidean. Then, we need to determine the temperature as a function of the layer number $\tau$. Let us denote the tensor rank as $m$ (see Fig.~\ref{fig3}). In VBS cases, $m$ is a small integer. The total quantum degree of freedom at a particular cut along $\tau$ axis is given by
\begin{eqnarray}
\chi = m^{\tau}. \label{chi}
\end{eqnarray}
Then we calculate the entanglement entropy. In 1D MPS cases, it has been known that the finite-entanglement entropy scaling is given by~\cite{Tagliacozzo,Pollmann,Huang,Nagy,Andersson}
\begin{eqnarray}
S_{FES} = \frac{c\kappa}{6}\ln\chi , \label{fes}
\end{eqnarray}
where $\kappa$ is the finite entanglement scaling exponent. There we have considered an infinitely-long open chain. Furthermore, the well-known Calacrese-Cardy formula~\cite{Holzhey,Calabrese} by CFT is given by
\begin{eqnarray}
S_{CFT} = \frac{c}{3}\ln\left(\frac{\beta}{\pi}\sinh\left(\frac{\pi L}{\beta}\right)\right) . \label{CC}
\end{eqnarray}
Substituting Eq.~(\ref{chi}) into Eq.~(\ref{fes}) and identifying $2S_{FES}$ with $S_{CFT}$, we obtain
\begin{eqnarray}
\tau\kappa\ln m = \ln\left(\frac{\beta}{\pi}\sinh\left(\frac{\pi L}{\beta}\right)\right) .
\end{eqnarray}
For high temperature cases, $S_{CFT}$ is approximately
\begin{eqnarray}
S_{CFT} \sim \frac{c}{6}\ln\left(\frac{\beta}{2\pi}\right) + \frac{c}{6}\frac{\pi L}{\beta} ,
\end{eqnarray}
and then we find
\begin{eqnarray}
T = \frac{\kappa\ln m}{\pi L}\tau .
\end{eqnarray}
Comparing it with $S_{FES}$, we find
\begin{eqnarray}
S_{FES} = \frac{c\kappa}{6}\tau\ln m = \frac{c\pi}{6}LT .
\end{eqnarray}
This shows the volume law of the entropy~\cite{Swingle}. The $\tau$ dependence on $T$ is different from that in MERA network.

\begin{figure}[htbp]
\begin{center}
\includegraphics[width=6cm]{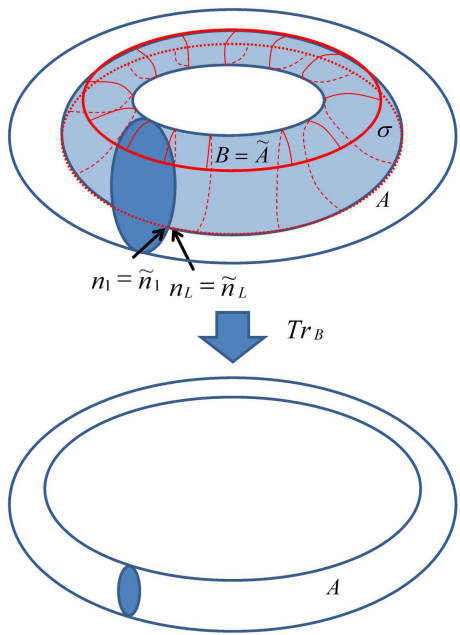}
\end{center}
\caption{Changing the order of taking boundary conditions.}
\label{fig4}
\end{figure}

Imagine that we change the order of taking the two boundary conditions as shown in Fig.~\ref{fig4}. We first pasted $n_{j}$ and $\tilde{n}_{j}$ together, and after that we take periodic boundary condition along the $x$ axis. Note that this may be somehow artificial since $A$ and $B$ are maximally entangled. Taking trace over $B$ degree of freedom in this case reproduces the original 1D quantum system under the periodic boundary condition (we do not use even MPS, just use the most general representation of $\left|\psi(\beta)\right>$). Then we observe the central charge $c$ in the original system. This is the most simple view.

We discuss implications of the present result and compare it with the thermal MERA network~\cite{Matsueda2,Matsueda3,Matsueda4}. Recently I have proved that the quantum information of the $d$-dimensional maximally entangled quantum state should be embedded into a $(d+1)$-dimensional classical space with hyperbolic metric~\cite{Matsueda6}. In string theory viewpoint, the maximally entangled state occurs in the $c\rightarrow\infty$ limit. The maximally entanglement is characterized by the entropy maximum ($\lambda_{i}=1/\chi$ for all $i$)
\begin{eqnarray}
S = -\sum_{i=1}^{\chi}\lambda_{i}\ln\lambda_{i} = \ln\chi.
\end{eqnarray}
On the other hand, in 1D cases, the finite-entanglement entropy scaling is given by
\begin{eqnarray}
S = \frac{1}{\sqrt{12/c}+1}\ln\chi ,
\end{eqnarray}
since $\chi=6/c(\sqrt{12/c}+1)$ in Eq.~(\ref{fes}). Actually, $S$ approaches $\ln\chi$ in the $c\rightarrow\infty$ limit. The is a classical limit of the tensor network space. This condition naturally leads to hyperbolic metric of the classical space~\cite{Matsueda6}. A related topic has been presented recently~\cite{Nozaki}. Now, we consider realistic spin-chain cases such as minimal series and Gaussian CFTs. Then, the metric would changes. In the following, we derive the explicite metric form, and discuss difference with MERA network.
In particular, we can obtain holographic eigenvalues representing geometric meaning of a correlation function, and those are useful for the discussion.

According to the previous examination~\cite{Matsueda6}, the metric is defined by using Fisher's information matrix as
\begin{eqnarray}
g_{\mu\nu} = q\sum_{i}\frac{1}{\lambda_{i}(z,x)}\frac{\partial\lambda_{i}(z,x)}{\partial x^{\mu}}\frac{\partial\lambda_{i}(z,x)}{\partial x^{\nu}}.
\end{eqnarray}
where $(x^{0},x^{1})=(z,x)$. Here, $q$ is a paramter representing curvature radius when the spacetime is curved. In the present case, this is only a uniform parameter. The important quantity is $\lambda_{i}(z,x)$ where $z$ and $x$ are 'hidden' or 'internal' parameters representing radial and spatial coodinates in the holographic space. We solve the differential equations when the left hand side is the flat metric. First we introduce the following decoupling:
\begin{eqnarray}
\lambda_{i}(z,x) = G(z,x)f_{i},
\end{eqnarray}
where $f_{i}$ is a monotonic function. Then the metric is given by
\begin{eqnarray}
g_{\mu\nu} &=& a\frac{1}{G(z,x)}\frac{\partial G(z,x)}{\partial x^{\mu}}\frac{\partial G(z,x)}{\partial x^{\nu}} , \\
a &=& q\sum_{i}f_{i} .
\end{eqnarray}
When $g_{zz}=g_{xx}=1$ and $g_{zx}=0$, we finally obtain the following asymptotic form in the 'UV' limit (in the previous examination the asymptotic form is obtained in the 'IR' limit):
\begin{eqnarray}
\lambda_{i}(z,x) = \frac{1}{4a}\left(z^{2}+x^{2}\right)f_{i} .
\end{eqnarray}
This represents how our target layer labeled by $z$ is correlated with the original 1D system. Since our tensor network is Euclidean, the triangular relation of the two-point distance is satisfied, and the distance characterizes the holographic spin correlation. The correlation reaches to infinitely long ranges, and thus the volume law of the entropy appears. This is in contrast to Ryu-Takayanagi formula for the holographic entanglement entropy in a context of AdS/CFT~\cite{Ryu}.

Summarizing, I have given a very simple proof for the correspondence between gapped 2D quantum systems and 1D CFT. For this proof, the finite-temperature tensor network formalism is a quite powerful tool. I argue that the present correspondence is similar to MERA but the metric of the holographic space is flat instead of hyperbolic. Our finding is a possible route to unify theoretical descriptions between gapped and critical systems. Furthermore, the compactification seems to be very different from bulk-to-edge correspondence, but we have found that the difference is only originated from the metric.

The author thanks Hosho Katsura, Jie Lou, and Naoki Kawashima for discussions.

\end{document}